\documentclass[twocolumn,showpacs,preprintnumbers,amsmath,amssymb,pra]{revtex4}
\usepackage{graphicx}
\begin{document}
\newcommand{\ud}{\mathrm{d}}

\title{Diffusion of triplet excitons in an
operational Organic Light Emitting Diode}

\author{M. Lebental$^{1}$, H. Choukri$^{1}$, S. Ch\'enais$^{1}$, S. Forget$^{1}$,
A. Siove$^{1}$, B. Geffroy$^{2}$, and E. Tuti\v s$^{3}$}
\affiliation{$^1$ CNRS, Universit\'e Paris Nord, UMR 7538,
Laboratoire
de Physique des Lasers, 93430 Villetaneuse, France\\
$^2$ CEA, Laboratoire de Physique des Interfaces et Couches Minces, Ecole Polytechnique,  91128 Palaiseau, France\\
$^3$ Institute of Physics, P. O. Box 304, HR-10000 Zagreb, Croatia}

\date{\today}

\begin{abstract}
Measurements of the diffusion length $L$  for  triplet excitons in
small molecular-weight organic semiconductors are commonly carried
out using a technique in which a phosphorescent-doped probe layer is
set in the vicinity of a supposed exciton generation zone. However,
analyses commonly used to retrieve $L$ ignore microcavity effects
that may induce a strong modulation of the emitted light as the
position of the exciton probe is shifted. The present paper
investigates in detail how this technique may be improved to obtain
more accurate results for $L$. The example of
4,4'-bis(carbazol-9-yl)1,1'-biphenyl (CBP) is taken, for which a
triplet diffusion length of $L$=16 $\pm$ 4  nm (at 3 mA/cm$^2$) is
inferred from experiments. The influence of triplet-triplet
annihilation, responsible for an apparent decrease of $L$ at high
current densities, is theoretically investigated, as well as the
'invasiveness' of the thin probe layer on the exciton distribution.
The interplay of microcavity effects and direct recombinations is
demonstrated experimentally with the archetypal trilayer structure
[N,N'-bis(naphthalen-1-yl)-N,N'-bis(phenyl)]-4,4'-diaminobiphenyl
(NPB)/CBP/ 2,9-dimethyl-4,7-diphenyl-1,10-phenanthroline (named
bathocuproine, BCP). It is shown that in this device holes do cross
the NPB/CBP junction, without the assistance of electrons and
despite the high energetic barrier imposed by the shift between the
HOMO levels. The use of the variable-thickness doped layer technique
in this case is then discussed. Finally, some guidelines are given
for improving the measure of the diffusion length of triplet
excitons in operational OLEDs, applicable to virtually any small
molecular-weight material.
\end{abstract}


\maketitle

\section{Introduction}

The performance of organic light emitting diodes (OLED) has been
pushed towards the ultimate limit of 100 \% internal quantum
efficiency thanks to the use of phosphorescent guest-host systems
\cite{baldo4,dandrade}. In phosphorescent organic light-emitting
devices, triplet exciton diffusion plays a major role: it has been
recently reported that one can take advantage of it to efficiently
monitor energy transport from the exciton creation zone up to the
emissive dopants to allow fine color tuning \cite{sun,sun2}. Exciton
diffusion also plays a key role in bilayer photovoltaic organic
devices as it governs the exciton dissociation efficiency
\cite{peumans,shao}.\\ In these cases, the interesting
characteristic of triplet excitons is their long diffusion length
compared to singlet excitons. Indeed, the diffusive properties of
triplet and singlet excitons are substantially different. While the
typical singlet exciton diffusion lengths are in the range of a few
nanometers in amorphous organic semiconductors \cite{choukri}, a
question arises about the order of magnitude for triplet excitons.
The diffusion length in a steady-state and linear regime is usually
described by $L_0=\sqrt{D\tau}$ , $D$ being the diffusion
coefficient and $\tau$ the exciton lifetime. On one hand, the
lifetime of triplet states (from $\mu$s to ms range) is much higher
than typical singlet states lifetime (ns). Meanwhile, it is not so
straightforward to compare the relative orders of magnitude of $D$
for singlets and triplets, since the physical mechanisms behind
their diffusion are fairly different. Exciton migration between two
non-emissive triplet states (e.g. host molecules in host-guest
phosphorescent systems) is a pure Dexter mechanism, consisting in a
simultaneous exchange of electrons in the LUMO and holes in the HOMO
levels. In contrast, energy transfer between two singlets can be
accounted both by a Dexter mechanism or a Forster non-radiative
dipole-dipole coupling even if the latter usually predominates
\cite{powell,forster}. As a consequence, $D$ coefficients could tend
to be lower for triplet excitons than for singlet excitons
\cite{baldo}.\\ It turns out to be an irrelevant task to seek an
universal order of magnitude for triplet exciton diffusion lengths,
which are expected to be highly dependant on the nature of the
material, its degree of purity, the nature and strength of the
excitation, etc. This difficulty is experimentally confirmed:
measured triplet diffusion lengths cover a range going from a few
nanometers in phosphorescent dendrimers \cite{namdas} to several
microns in pure organic crystals \cite{ern}. Furthermore, even for a
well-known material such as CBP, the reported diffusion lengths are
highly scattered
\cite{zhou,giebink,dandrade2,matsusue}.\\
It is then of foremost importance to develop both theoretical and
experimental tools to improve our understanding of triplet exciton
dynamics. Triplet migration has been a topic of intense research
firstly in organic crystals \cite{ern} and aromatic hydrocarbons
\cite{levshin} and then more recently in amorphous organic
semiconductors \cite{rothe,devi,fisshuk}. Along with theoretical
work, it is essential to have reliable direct measurements of
diffusion lengths to support both theory and device design. The aim
of the present paper is to identify the main physical processes and
parameters that have to be taken into account to perform a
meaningful measurement of triplet exciton diffusion in an
operational
device, and to propose some guidelines for extracting these parameters experimentally.\\
The paper is organized as follows: In Section
\ref{sec:structure-optimisee}, we present an experimental
measurement of the triplet diffusion length in CBP based on the
technique first proposed by Baldo et al. \cite{baldo3}. To enhance
the reliability of the measurement, we used a device specially
designed to exclude two physical effects likely to mask the
diffusion process, namely \emph{microcavity effects} and \emph{bulk
carrier recombinations}. Indeed, the optical field variation related
to microcavity effects is huge this case (the technique requires
thick diodes), even if its influence is usually neglected in
comparable devices \cite{sun}, leading to questionable values for
the diffusion lengths. These experimental results serve as a basis
for a discussion of the relevant key points in the following
sections. Section \ref{sec:diff-th} is a theoretical investigation
which aims at precisely defining the exciton diffusion length, gives
the analytical solution for the steady-state exciton distribution in
presence of triplet-triplet annihilation, and proposes a
quantitative criterium to quantify the strength of this bi-particle
process. The influence of the thin sensing layer is also
investigated, motivated by an insight that its presence may
considerably alter the distribution of triplets in the device. In
Section \ref{sec:artefacts}, we use an archetypal diode structure
\cite{dandrade2,sun} to experimentally illustrate the combined
influence of bulk carrier recombinations and optical field
variations. For the same device we also demonstrate the poor
hole-blocking efficiency of a hetero-junction usually considered as
a strong barrier for holes and therefore as the exciton generation
zone \cite{sun,zhou}. We conclude by the comparison of our result
for the triplet diffusion length in CBP to the values published by
other authors. The concluding section lists our recommendations for
a reliable measurement of triplet diffusion lengths.

\section{Diffusion measurement}\label{sec:structure-optimisee}

\subsection{Choice of the technique}

Two types of methods have been commonly used to measure the
diffusion length of triplet excitons and both have been applied to
CBP: excitons are created either by optical excitation or by carrier
recombination in an operational OLED device. The techniques based on
optical excitation enable easy time-resolved studies, while those
based on electrical excitation provide a higher control of the
exciton formation zone and are closer to the operating conditions of
real devices. With optical techniques (photocurrent spectroscopy
\cite{yang,matsusue} or time-resolved spectral decay analysis
\cite{giebink}), obtaining a clear signature of diffusion is
intricate when the absorption length of the laser (of the order of
50 nm \cite{giebink}) has the same order of magnitude as the
diffusion length, which is the case in practice. This issue
disappears under electrical excitation, where localized "sheets" of
excitons are achievable. In fact the strong localization of the
exciton formation is a very robust consequence of the energy barrier
and the carrier accumulation at the heterojunction in multilayer
OLEDs. The spatial scale of the exciton formation zone is then of
the order of the thickness of a few molecular monolayers, reflecting
the space distribution of blocked carriers. This is much less than
one is likely to get under most favorable circumstances from the
space charge effects caused by the low carrier mobility, as
illustrated in Ref. \cite{tutisFJ}. Thus in general, exciton
formation zones are highly localized at hetero-junctions
\cite{baldo2,dandrade2,zhou}. Furthermore all quenching processes
are included, like exciton-exciton or exciton-polaron annihilation,
which may be desirable when one attempts to obtain an effective
diffusion length directly exploitable to build real light-emitting devices.\\
In the technique used by D'Andrade et al. \cite{dandrade2}, excitons
are created at one edge of a thick CBP layer doped with Iridium(III)
tris(2-phenyl-pyridinato-N,C$^2$') (Ir(ppy)$_3$), and the light
emitted by the phosphor is collected for various thicknesses of the
doped layer. A diffusion length of 8.3 $\pm$ 1 nm was then derived
for CBP. Zhou et al. \cite{zhou} pointed out that these measurements
yielded information about an Ir(ppy)$_3$-doped CBP system rather
than about a pure undoped CBP layer. They proposed a refined model
to extract the diffusion length in pure CBP from the same set of
data and obtained about 60 nm. However, in their fitting procedure
two unknown parameters have to be extracted simultaneously (namely
the diffusion constants for the doped and the undoped region) under
the assumption that all excitons are created at the interface
between the Hole Transporting Layer (HTL) and the Emitting Layer
(EML) (in this case between NPB and CBP). As shown in Sec.
\ref{sec:npb/cbp}, this assumption is rather questionable for
this particular heterojunction. \\
In the present paper, we use the technique described by Baldo et al.
\cite{baldo3} in which excitons are generated at a heterojunction,
diffuse in a neat undoped region until they reach a thin
phosphorescent layer acting as a probe. Since only short-range
($\sim$1 nm) Dexter transfer is possible from host to host or from
host to guest in the case of triplets, the emissive layer truly acts
as a local probe for triplet excitons, with a spatial resolution
almost only limited by the thickness of the sensing layer.\\

\subsection{Choice of parameters}

\begin{figure}
\begin{center}
\includegraphics[width=1\linewidth]{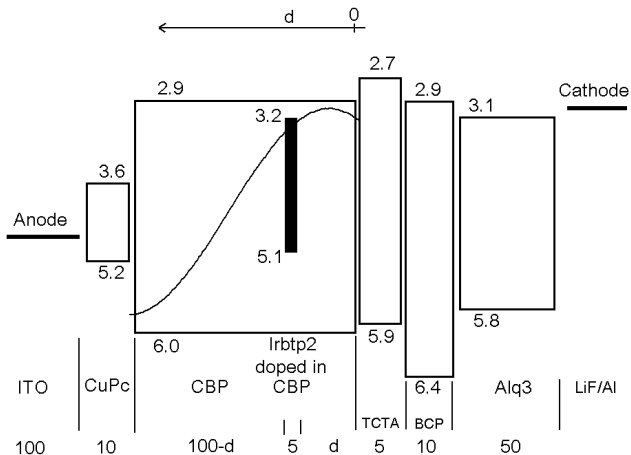}
\end{center}
\caption{Structure of the diode used in Sec.
\ref{sec:structure-optimisee}. The layer thicknesses are indicated
on the bottom in nanometers. The HOMO and LUMO energy levels are
taken from literature \cite{hill,ikai} and specified for each
compound (negative values). The profile of the optical field is
plotted in the CBP layer (see Sec. \ref{sec:champoptique} for
details).}
    \label{fig:diodediffusion}
\end{figure}

The choice of CBP in this study is motivated by the fact that it has
been the subject of many investigations, leading to several
measurements of the triplet exciton diffusion length, with different
techniques and highly scattered experimental results.\\
The OLED structure and the HOMO/LUMO levels of the different
materials are shown in Fig. \ref{fig:diodediffusion}. It is made up
of a standard layer stack embedding the CBP matrix: indium tin oxide
(ITO) anode coupled to a copper phtallocyanine (CuPc) layer for hole
injection (HIL), aluminium tris(8-quinolinolato)(Alq$_3$) layer for
electron transport (ETL) and LiF/Al cathode for electron injection.
In order to efficiently block both electrons and holes and generate
excitons in the form of a localized "sheet", two additional layers
were used, each of them being well-known to efficiently block one
type of carriers: 4,4',4''-tris(carbazol-9-yl)-triphenylamine (TCTA)
for electrons and BCP for holes. The triplet exciton sensor (the
emitting layer EML) is a 5-nm-thick layer of CBP doped with 6 \%
Iridium(III) bis[2-(2'-benzothienyl)-pyridinato-N,C3']
acetylacetonate (Ir(btp)$_2$ acac) and inserted at a position
defined by the $d$ parameter equal to the distance between the
CBP/TCTA interface and the center of the sensing layer (see Fig.
\ref{fig:diodediffusion}). The choice of Ir(btp)$_2$ was motivated
by its phosphorescent efficiency \cite{tsuboi,adachi} and its
emission spectrum easily resolved from those of the other compounds
used in this OLED
\cite{lamansky} (see on Fig. \ref{fig:pl}).\\
The thicknesses of the different layers are given in Fig.
\ref{fig:diodediffusion}. The CBP layer is noticeably thick because
relatively long diffusion lengths are expected. Moreover the Alq$_3$
and CBP layers are optimized so that the generation zone of excitons
is located at a position where the optical field corresponding to
the red emission of the phosphorescent layer is as flat as possible
over a long distance. If this condition is not met, the variation of
the optical field should be carefully taken into account and it may
be difficult to decouple it from the effect of diffusion, as further
discussed in Sec. \ref{sec:artefacts} where details about the
calculation of the optical field are also given. On the relevant
scale (i.e. $d$ between 0 and 40 nm), its variation appears to be
less than 20 \% and could hardly be reduced. The thickness of the
EML has been set to 5 nm. This is thin enough to limit the influence
of its position $d$ on the optical field. A theoretical
investigation of the role of the sensing layer in electronic
properties and exciton transport will be
exposed in Sec. \ref{sec:diff-th}.\\
When evaporating the thin sensing layer, the question may arise of
how the doping rate and the layer thickness can be accurately
controlled and reproduced \cite{zhou}. We solved this issue by
systematically making four devices in each single run: two of them
corresponding to a "reference diode" (fixed $d$ parameter), and the
other two to some other value of $d$. The electroluminescence
was then always normalized to the reference diode.\\
The glass substrate covered by ITO was cleaned by sonication and
prepared by a UV-ozone treatment. The layers were then deposited by
sublimation under high vacuum (10$^{-6}$- 10$^{-7}$ mbar) at a rate
of 0.1-0.2 nm/s in a thermal evaporator. An \emph{in situ} quartz
crystal was used to monitor the thickness of the layer depositions
with a precision of 5 \%. The organic materials and the LiF/Al
cathode were deposited in a one-step process without breaking the
vacuum.\\
After deposition, all the measurements were performed at
room temperature and under ambient atmosphere, without any
encapsulation. For each diode with a specific position $d$ of the
thin Ir(btp)$_2$:CBP layer, the current-voltage-luminance
characteristics and electroluminescence spectra (in the direction
normal to the substrate) were collected and recorded with a PR 650
SpectraScan spectrophotometer for different currents from 0 to 50
mA, corresponding to a current density $J$=0-166 mA/cm$^2$.

\subsection{Results}

\begin{figure}
\begin{center}
\includegraphics[width=1\linewidth]{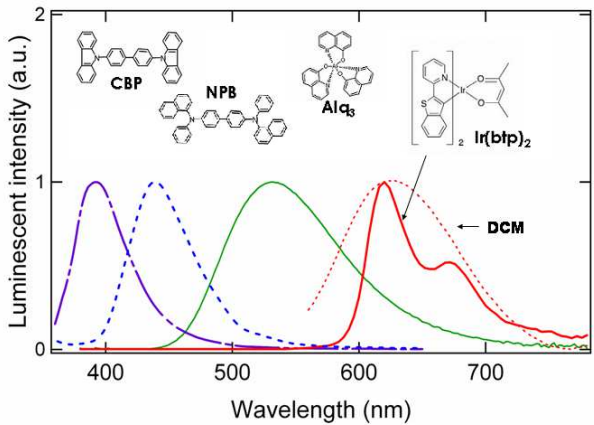}
\end{center}
\caption{Photoluminescent spectra of some of the compounds used in
this paper with their chemical structure. The electroluminescent
spectrum of DCM is plotted in dotted line.}
    \label{fig:pl}
\end{figure}

\begin{figure}
\begin{center}
\includegraphics[width=0.9\linewidth]{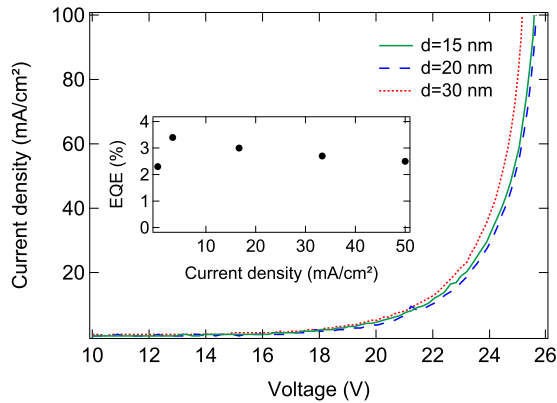}
\end{center}
\caption{Current density versus voltage curves for diodes with
different $d$ positions of the probe layer. Insert: External quantum
efficiency (EQE) versus current density for a diode with $d$=7.5
nm.}
    \label{fig:iv}
\end{figure}

\begin{figure}
\begin{center}
\includegraphics[width=1\linewidth]{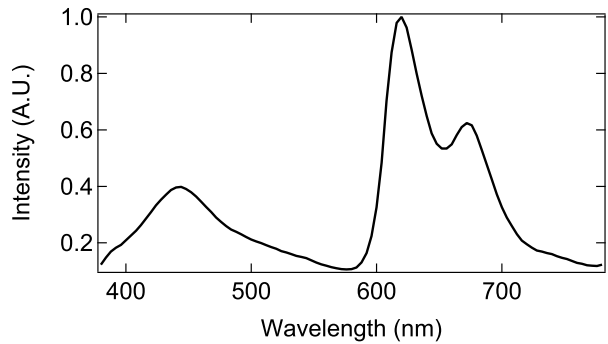}
\end{center}
\caption{Electroluminescent spectrum for a diode described in Fig.
\ref{fig:diodediffusion} with $d$=12.5 nm at a current density of
3.3 mA/cm$^2$.}
    \label{fig:spectre-opt}
\end{figure}

The Current-Voltage (I-V) curves appear in Fig. \ref{fig:iv}. The
high voltage threshold (about 20 V) is consistent with the unusually
large thicknesses of our devices compared to those commonly reported
in the literature. It may be noticed that the I-V characteristics
are similar whatever the position $d$ of the sensing layer, showing
its negligible influence on the transport properties.\\
The external quantum efficiency is maximum for a current density
around 3 mA/cm$^2$ with a value of 3.2 $\pm$ 0.5 \% for $d$ = 7.5 nm
(see insert of Fig. \ref{fig:iv}), and we observed a roll-off of the
external quantum efficiency with the current, which is a classical
feature of phosphorescence-based OLEDs usually attributed to
triplet-triplet quenching
between guest molecules \cite{baldo2,reineke}.\\
The typical electroluminescence spectrum shown in Fig.
\ref{fig:spectre-opt} comprises two different contributions. The red
structured peak is the clear signature of Ir(btp)$_2$ (see Fig.
\ref{fig:pl}), but the blue peak (centered at 450 nm) cannot be
associated with any photoluminescence (PL) spectrum. Furthermore,
the optical field variations do not allow explaining such a
difference between the observed electroluminescence spectrum and any
of the PL spectra. We then conclude that this blue peak originates
from exciplexes \cite{weller} formed at the TCTA/BCP interface,
which is consistent with the large energy shifts between their LUMO
(0.4 eV) and HOMO levels (0.5 eV) \cite{cocchi,li} and with a clear
spectral redshift with respect to the TCTA PL spectrum (maximum
around 410 nm) \cite{shirota} \footnote{A rough estimate of the
exciplex energy, notwithstanding the exciplex binding energy and
energetic disorder, is given by the difference between the TCTA HOMO
level (-5.9 eV) and the BCP LUMO
level (-2.9 eV) and corresponds to a wavelength of about 410~nm.}.\\
For distances $d$ long enough for singlet exciton density to vanish
(typically less than 10 nm \cite{choukri}), the red emission from
Ir(btp)$_2$ may have several origins: it may result from direct
recombinations between holes and electrons traveling in bulk CBP
(the electrons would have crossed the thin TCTA layer by tunneling
or any other process); it may also come from triplet excitons
diffusing from the CBP/TCTA interface. The mechanism leading to
triplet excitons being formed in CBP from exciplexes is beyond the
scope of this paper and deserves further investigation since many
processes can be invoked. Whatever the mechanism at work, the net
result is a triplet exciton population in a restricted area around
the CBP/TCTA heterojunction, which can not come back to the TCTA
layer due to its
larger energy gap. \\

\begin{figure}
\begin{center}
\includegraphics[width=1\linewidth]{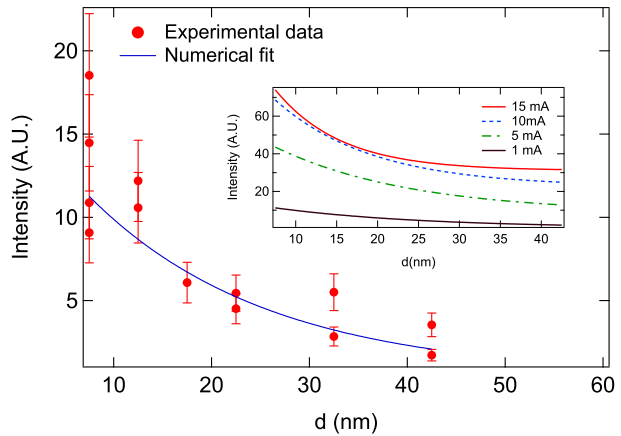}
\end{center}
\caption{The circles correspond to the red detected intensity
corrected by the optical field versus the $d$ position of the probe
layer at $J=3.3$ mA/cm$^2$. The fit by Eq. (\ref{eq:diff-simple}) is
plotted in continuous line. Insert: Fits from Eq.
(\ref{eq:diff-simple}) of experimental data for different current
densities.}
    \label{fig:resultat-opt}
\end{figure}

\begin{figure}
\begin{center}
\includegraphics[width=0.65\linewidth]{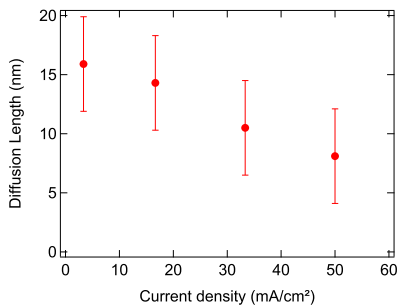}
\end{center}
\caption{Effective diffusion length $L$ of triplet exciton versus
the current density, defined from Eq. (\ref{eq:diff-simple}) and
inferred from the fits shown in the insert of Fig.
\ref{fig:resultat-opt}.}
    \label{fig:long-eff}
\end{figure}

The intensity emitted by the phosphorescent material for a given
current density $J$=3.3 mA/cm$^2$ is integrated over its spectral
range and then plotted versus $d$ in Fig. \ref{fig:resultat-opt}
after dividing by the amplitude of the optical field at the same
position. A downward trend from the interface ($d$=0) is clearly
visible, which evidences a diffusion process, since direct
recombinations should yield a signal which is independent of $d$
(after compensation of the optical field variations). When the
current is increased, the probe intensity shows a kind of "plateau"
for large $d$ values, which can be attributed to direct
recombinations and indicates that more and more electrons pass
through the thin TCTA layer. The experimental data can consequently
be fitted by the following expression:
\begin{equation}\label{eq:diff-simple}
I(d,J)=[A(J)+B(J)\,e^{\,-d/L(J)}]\times E_{opt}(d)
\end{equation}
where $I(d,J)$ is the intensity of light emitted for a position $d$
of the sensing layer at a given current density $J$, and
$E_{opt}(d)$ is the optical field. Here, we fit the data with a
constant ($A$) and a simple exponential decay ($B\,e^{-\,d/L}$),
which is much simpler than the refined analysis presented in Sec.
\ref{sec:diff-th}. However it still gives the typical distance $L$
(called hereafter "effective diffusion length of triplet excitons")
over which excitons can diffuse, even if the non-linear contribution
of the triplet-triplet annihilation does not allow, strictly
speaking, to consider an exponential (or bi-exponential) behavior.
The constant term $A$ stands for direct recombinations so that $B/A$
corresponds to the ratio of light intensity generated from triplet
excitons over light intensity generated from direct recombinations.
$A$, $B$ and
$L$ depend \emph{a priori} on the current density $J$.\\
The fits using Eq. (\ref{eq:diff-simple}) and presented in insert of
Fig. \ref{fig:resultat-opt} show that the number of direct
recombinations grows with the current density as expected. From the
fits, the effective diffusion length of triplet excitons is
estimated to be 16 $\pm$ 4 nm at low currents, and turns out to
decrease down to 8 nm when the current increases (see Fig.
\ref{fig:long-eff}) as expected due to bimolecular interactions such
as triplet-triplet annihilations or
triplet-polaron quenching \cite{baldo,baldo2,reineke}.\\
This effective diffusion length, even at low current densities, is
small compared to what could be expected from the high lifetime of
CBP triplet excitons (14 $\pm$ 8 ms \cite{giebink}). Moreover it is
lower than previously reported values under electrical excitation:
46 $\pm$ 3 nm at 10 mA/cm$^2$ by Sun et al. \cite{sun} and 60 nm at
unspecified current density by Zhou et al. \cite{zhou}. However we
discuss in Sec. \ref{sec:artefacts} how direct recombinations,
microcavity effects and barrier energetics may invalidate some
aspects of these measurements.\\
In order to unambiguously attribute the downward tendency to triplet
exciton diffusion, a control experiment was carried out with a
fluorescent compound,
4-(dicyanomethylene)-2-methyl-6-(p-dimethylaminostyryl)-4H-pyran
(DCM), instead of the phosphorescent Ir(btp)$_2$. Actually DCM
triplet states do not emit light and CBP triplet excitons cannot
transfer their energy towards singlet states of DCM as Dexter
transfer requires total spin conservation
\footnote{Triplet-to-singlet transfer might be possible if the donor
exciton breaks up and reforms on the acceptor via incoherent
electron exchange. However, as pointed out by Baldo et al.
\cite{baldo3}, this has to be considered as very unlikely since the
energy required for dissociation, i.e. the exciton binding energy,
approaches 1 eV in most systems.}. Therefore the comparison between
both devices allows a clear distinction between triplet diffusion
and direct recombinations. Moreover the DCM and Ir(btp)$_2$ emission
spectra exhibit a similar envelope (see Fig. \ref{fig:pl}) so that
the optical field effect is not modified. The only difference with
respect to previous experiments is the lower doping rate of DCM in
the CBP matrix (1.5 \% in weight), necessary to limit concentration
quenching \footnote{The emitted intensity from DCM and Ir(btp)$_2$
can not be directly compared to each other due to their different
quantum yields and concentration.}. As a result, OLEDs were realized
with the DCM-doped CBP thin layer set at two $d$ positions
characterized by an optical field having almost identical values
($d$=12.5 nm and $d$=42.5 nm), and which are both far enough from
the recombination zone to neglect the influence of singlet exciton
diffusion. The measured DCM emitted intensity is 15 \% smaller at
$d$=12.5 nm than at $d$=42.5 nm ($J$ = 3.3 mA/cm$^2$) which is
compatible with the measurement uncertainties and evidences direct
recombinations in bulk CBP. In the case of the Ir(btp)$_2$:CBP doped
layer, the difference between the intensity emitted at these two
positions is much larger (5 times more red light at $d$=12.5nm than
at $d$=42.5 nm), which is then an
unambiguous signature of triplet exciton diffusion.\\
Moreover, with the DCM:CBP layer, we clearly observed that the color
emitted by the diode shifted from the sky-blue emission of dominant
exciplexes at low currents to magenta as the current was increased,
which is an additional proof that the "plateau" observed above
(corresponding to the $A(J)$ parameter in Eq.
(\ref{eq:diff-simple})) is the manifestation of current-dependent
direct recombinations.\\
The strategy developed in this Section to obtain an estimation of
the diffusion length, albeit specific to CBP, can be applied to
virtually any material. It is the objective of the following
Sections to discuss in more detail the physical parameters
influencing the measurement of triplet exciton diffusion lengths. We
need first to examine, from a theoretical point of view, how the
intensity emitted by the probe relates to the actual exciton
density, especially when triplet-triplet annihilation is present
or/and when the probe layer itself affects the exciton motion.

\section{Diffusion length}\label{sec:diff-th}

In the limit of low exciton concentration, their distribution in
space $n(x)$ goes exponentially with $x$, the distance from the
source, with a scale set by the intrinsic diffusion length of
triplet excitons $L_0$. However, the distribution in space is not
expected to be the same in the case when the triplet-triplet
quenching activates, as the strength of the source of excitons
increases. Additionally, if a probe used for detecting triplet
excitons is efficient in their trapping and recombining, it may
considerably disturb their distribution in space. These effects are
analyzed below starting from the usual diffusion-decay model. The
dependence of the signal on the distance from the source is derived
when one or both effects are present. It is concluded that the
invasiveness of the probe does not present a serious obstacle in
extracting the proper value of $L_0$ from the experiment.
Conversely, the effect of triplet-triplet quenching at higher source
intensity, if not analyzed properly, may lead to significant
underestimate of $L_0$.

\subsection{Fundamental equation}

The equation that governs the diffusion of triplet excitons in an
organic one-dimensional \footnote{In Sec.
\ref{sec:structure-optimisee}, the triplet excitons diffuse from the
CBP/TCTA hetero-junction (plane $x=0$) with a $G$ constant rate and
an uniform distribution in the $y$ and $z$ directions. This
three-dimensional (3D) problem can thus be solved as an
one-dimensional problem thanks to the planar geometry of the source
term, but the parameters such as the diffusion length and the
diffusion coefficient are still the same (ie. defined in 3D). In
this configuration, the diffusion length is the same in the real
three-dimensional system or in its reduction along the $x$ axis.
Thus we will only consider the latter case.} layer
is\begin{equation} \frac{\partial n}{\partial
t}=D\frac{\partial^{2}n}{\partial
x^{2}}-\gamma_{T}\,n-\gamma_{TT}\,n^{2}\label{eq:DiffTime}\end{equation}
The right hand side of this equation is made up of three different
contributions. First the term $D\frac{\partial^{2}n}{\partial
x^{2}}$ characterizes the genuine diffusion of triplet excitons
through the $D$ diffusion coefficient, which is assumed to be
isotropic and constant through the whole CBP layer. Then, the term
$\gamma_T~n$ gathers all the processes responsible for a decreasing
of the triplet exciton density, where only one triplet exciton is
involved in. It is made up of the radiative and non-radiative
desexcitations $\gamma~n$, the singlet-triplet annihilation
$\gamma_{ST}~n_S~n$, the polaron-triplet annihilation
$\gamma_{PT}~n_P~n$ \cite{reineke}, and eventually of other
quenching processes. For the sake of simplicity, we assume an
uniform value of $\gamma_T$, which is probable since the densities
of singlet excitons and polarons become roughly constant (or even
negligible) in a bulk material a few nanometers away from the
generation zone of excitons. Finally the term $\gamma_{TT}~n^2$
corresponds to the triplet-triplet annihilation which was observed
under typical OLED operation conditions \cite{baldo2,reineke}. Its
influence will
be evaluated below.\\
For fitting with experiments presented in Sec.
\ref{sec:structure-optimisee}, we focus on the stationary state
solution of equation, \begin{equation}
D\frac{\partial^{2}n}{\partial
x^{2}}-\gamma_{T}n-\gamma_{TT}n^{2}=0,\label{eq:DiffST}\end{equation}
in the presence of a steady source at $x=0.$ The solution $n(x)$ is
sought in the portion of the space $x>0$, away from the source. The
strength of the source, $G,$ sets the value of the exciton current
at the $x=0$ boundary, $G=-Dn'(0).$\\
Obviously, the relative importance of the two decay terms in Eq.
(\ref{eq:DiffST}) depends on the concentration of excitons, with
their influence being comparable at the $n_{0}$ characteristic
density: $n_{0}\equiv\gamma_{T}/\gamma_{TT}.$ In the limit of the
rare exciton gas, $n\ll n_{0}$, the distribution of triplet excitons
is given by the simple exponential dependence,
$n(x)\propto\exp(-x/L_0),$ set by the only intrinsic length scale in
the equation, the \emph{triplet diffusion length},\begin{equation}
L_0\equiv\sqrt{\frac{D}{\gamma_{T}}}.\label{eq:dT-def}\end{equation}

\subsection{Triplet-triplet quenching}

\begin{figure}
\begin{center}
\includegraphics[width=1\linewidth]{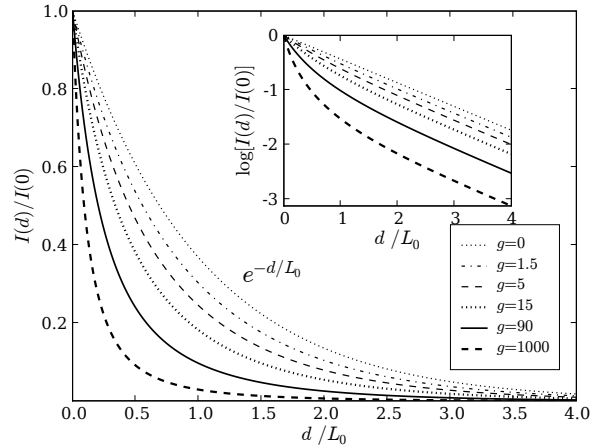}
\end{center}
\caption{The calculated dependence of the emitted signal in the
presence of triplet-triplet quenching and non-invasive probe
inferred from Eq. \ref{eq:nx-TT-ni-Id}. The insert gives the
semi-logarithmic view of the same set of data.}
    \label{fig:TTni}
\end{figure}

The spatial dependence $n(x)$ complicates
as the strength of the source $G$ increases, and when the only dimensionless
parameter of the problem,
\begin{equation}
g\equiv\frac{G\,\gamma_{TT}}{L_0\,\gamma_T^2},\label{eq:g-def}
\end{equation}
rises above unity \footnote{The quenching among triplet excitons in
host is expected when they come close in terms of the Dexter radius,
and similarly the quenching of triplet excitons and holes
(polarons). This is since otherwise it is difficult to make the
spin-flip required for triplet excitons to decay. This implies
$\gamma_{TT}$ and $\gamma_{PT}$ being of the same order of
magnitude. Therefore frequency of two processes is essentially given
by the ratio of the concentrations of triplets and polarons
$\gamma_{TP}n_Pn_{T}/\gamma_{TT}n_{T}^{2}\sim n_P/n_{T}.$ For
typical values of current density, mobility, electric field, and
molecular size ($\mu_{p}\sim10^{-4}$cm$^{2}$/Vs, $F\sim1$ MV/cm,
$J\sim100$ mA/cm$^{2}$, $a\sim0.6$ nm) the concentration of polarons
is of the order of $n_P\sim10^{-7}$ per molecule, while the creation
rate of excitons/exciplexes at the heterojunction, for the same set
of parameters, is estimated to $\nu\sim10^{3}s^{-1}$ per molecular
cross section. The density of triplets $n_{T}$ is obtained upon
considering the fraction $\alpha$ of excitons created being
triplets, considering their diffusion over length $L_{0}$, as well
as accounting for their decay rate $\gamma_{T}$. For
$\gamma_{T}\sim10^{6}$s$^{-1}$ and $L_{0}\sim50\,a$, the
concentration of triplets per molecule turn to be
$n_{T}\sim\alpha\times10^{-4}$. This gives a plenty of room for the
possibility that $n_{T}\gg n_P,$ corresponding to the situation
where triplet-triplet quenching is much more frequent than the
quenching of triplets on polarons.}. The nonlinear differential
equation Eq. (\ref{eq:DiffST}) for the steady state has to be solved
in that case, and the details are given in Appendix
\ref{sec:annexe-TT}. The solution may be written in the form
\begin{equation}
n(x)=\frac{3\,n_{0}}{2\left(\sinh\frac{x+d_{TT}}{2L_0}\right)^{2}},
\label{eq:nx-TT-ni-sinh}
\end{equation}
where a new length scale $d_{TT}$ is introduced,
$d_{TT}/L_0=\ln(1/\mu),$ with the parameter $\mu$ related to the
dimensionless source strength $g$ through
\begin{equation}
g=\frac{6\left(1+\mu\right)\mu}{\left(1-\mu\right)^{3}}.\label{eq:g-mu}
\end{equation}
The parameter $\mu$ approaches zero for a weak source $g\ll1$, and
approaches unity for a strong source, $g\gg1.$ A strong source
implies $d_{TT}\ll L_0,$ with $n(x)$ falling much faster than
$\exp(-x/L_0)$ in the region $0<x<L_0$. The profile of $n(x)/n(0)$
determines the variation of the emitted light in the limit of a
non-invasive and infinitely thin probe layer at the $d$ position,
\begin{equation}
\frac{I(d)}{I(0)}=\frac{\left(\sinh\frac{d_{TT}}{2L_0}\right)^{2}}
{\left(\sinh\frac{d+d_{TT}}{2L_0}\right)^{2}}.
\label{eq:nx-TT-ni-Id}
\end{equation}
Examples of this dependence for various values of the $g$ parameter
are shown in Fig. \ref{fig:TTni}. The figure shows that the spatial
decay rate effectively increases as the strength of the exciton
source increases, in accordance with our experimental findings.
However, as the variation of intensity with $d$ is no more
exponential when the influence of triplet-triplet quenching grows,
the concept of "effective diffusion length", largely used in the
literature as well as in Sec. \ref{sec:structure-optimisee} is,
strictly speaking, an ill-defined parameter. In ideal cases, the
data should enable by a proper fit the extraction of $L_0$ and
$d_{TT}$, from which it should be possible to derive $\gamma_{TT}$.
In our case however, there are several obstacles against extracting
the value of the triplet-triplet quenching parameter $\gamma_{TT}$
from the experimental data. First, less scattered experimental data
may be required for precise evaluation of the parameter $g$
determining the shape of the curves in Fig. \ref{fig:TTni}. More
importantly, in order to determine $\gamma_{TT}$ from $g$ one still
has to know other parameters in Eq. (\ref{eq:g-def}), most notably
the strength of the exciton source
$G$, and calculating its absolute value is not straightforward.\\
However, it is informative to have an idea of the order of magnitude
that should be expected for $g$ in real cases. For that purpose, we
can consider that a given ratio $\eta$ of injected carriers are
converted into potentially diffusing triplet excitons. Then at low
current density, $\eta$ would reach the maximum value of 0.75 if
neither exciplex would be involved in, nor other lossy intermediate
states. The strength of the exciton source $G$ can thus be written
$G\simeq\eta\,J/q$, where $q$ is the elementary charge. The other
parameters needed to compute $g$ are taken from \cite{giebink}
($\tau_T$ = 1/$\gamma_T$ = 15 ms and $\gamma_{TT}$ $\simeq$
1.10$^{-14}$ cm$^3$/s), and our experimental value of $L_0$ is
considered ($L_0$ $\sim$ 15 nm). Then for $J\simeq$ 1 mA/cm$^2$ and
$\eta=0.75$, $g$ is about 1000. Whatever the realistic value of
$\eta$, $g$ is probably higher than 1 \footnote{A rough estimation
of $\eta\sim 10^{-2}$ can be inferred from the red part of the
experimental spectra (see Sec. \ref{sec:structure-optimisee}),
assuming 20 \% efficiency in extracting photons and an isotropic
emission over the upper half-plane. For $J$=1 mA/cm$^2$, this
implies $g\sim10$.}. The triplet-triplet annihilation is then
supposed to have a strong influence on measurements even at low
current densities, which may be the explanation of such small
effective diffusion lengths reported in electrically-driven
structures.

\subsection{Influence of the sensing layer}\label{sec:th-layer}

The detection of the exciton diffusion by the probe technique used
in Sec. \ref{sec:structure-optimisee} assumes that the sensing layer
at different position $d$ does not influence the conditions of
exciton generation, diffusion and recombination. The 'invasiveness'
of the sensing layer is, however, not fully avoidable in practice.
Two aspects of the invasiveness are imaginable. One aspect is
related to the trapping of charge carriers in the sensing layer,
whereas the other relates to the trapping of excitons.\\
The change of the electric field distribution caused by trapped
charges \cite{campbell}  is not expected to affect very much the
motion of excitons which are neutral objects.  On the other hand it
may be noted that, for a given external voltage, the difference in
the spatial profile of the electric field in two devices with
different position $d$ of the sensing layer implies different charge
distribution among interfaces, with a probable effect on exciton
generation and recombination. In our case however, as shown in Fig.
\ref{fig:iv}, this effect was negligible. In a more general fashion,
it can be ignored as soon as the devices are compared for the same
value of the current running through them.\\
If the effects of charge trapping do not seem relevant in the
experiments of Sec. \ref{sec:structure-optimisee}, the consequences
of exciton trapping must be considered more carefully. Actually a
sensing layer of thickness $\delta=5$ nm is introduced to absorb a
fraction of the triplet excitons and convert them into photons.
Ideally, this layer is thin and absorbs only a small fraction of
triplets, without significantly perturbing their distribution in
space. In practice, this implies rather small signal from the probe,
which may then be masked by the light emission from exciton
recombinations elsewhere in the device. Realistically, the doping of
the order of a few percents made through several monolayers of the
host material, may already represent rather invasive probe. Actually
the separation between doped molecules is of the order of few
molecular sizes, and then also of the order of the Dexter radius,
the scale involved in the diffusion of triplet excitons. The sensing
layer then disturbs the genuine dependence $n(x)$, implicating a
deviation from the ideal case. This effect may be easily modeled by
replacing the decay term $-\gamma_{T}\,n$ by as stronger one
$-\gamma_{SL}\,n$, with $\gamma_{SL}>\gamma_{T}$ within the sensing
layer $d<x<d+\delta.$ The measured quantity is then the number of
excitons absorbed by the sensing layer per unit time, proportional
to\begin{equation} I(d)=\int_{d}^{d+\delta}\gamma_{SL}\,n(x)\,\ud
x.\label{eq:I-of-d-def}\end{equation}

\begin{figure}
\begin{center}
\includegraphics[width=1\columnwidth]{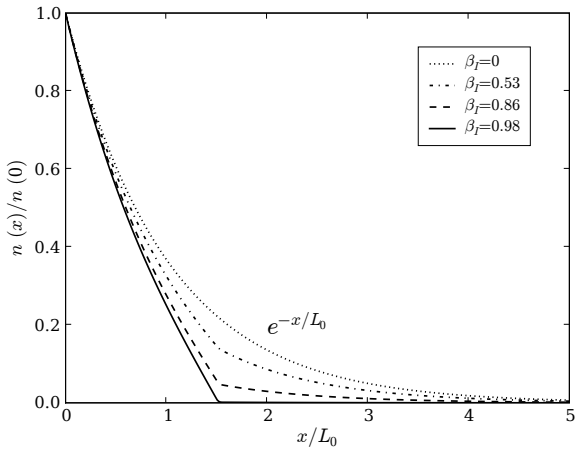}
\end{center}
\caption{The influence of the probe on the distribution of excitons
in space $n(x)/n(0)$ for various levels of invasiveness $\beta_I$.
The sensing layer of thickness \emph{$\delta=0.05\, L_0$} is placed
as \emph{$d=1.5\, L_0.$}}
    \label{fig:nx-si}
\end{figure}

\begin{figure}
\begin{center}
\includegraphics[width=1\columnwidth]{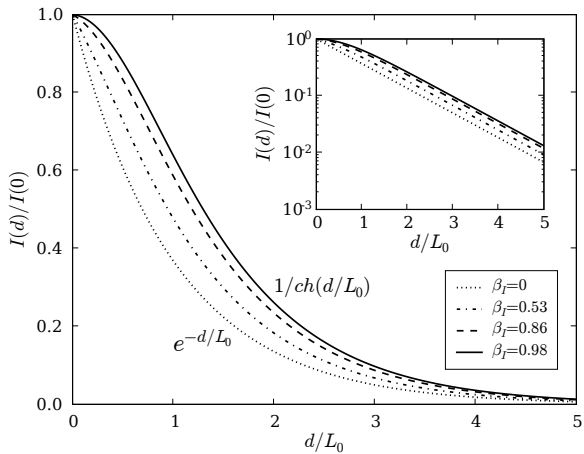}
\end{center}
\caption{The dependence of the signal intensity $I(d)/I(0)$ on the
position $d$ of the probe, for various levels of its invasiveness
$\beta_I$. The formulae are indicated on the graph for the limits of
non-invasive and strongly invasive probe. }
    \label{fig:I-si}
\end{figure}

For the sake of simplicity, the effect of the invasiveness of the
sensing layer is first examined in the limit of low exciton density,
when the triplet-triplet quenching term may be neglected, but in
fact it does not modify strongly these results (the full problem
including triplet-triplet quenching is exposed in Appendix
\ref{sec:annexe-semi-invasive-and-TT}). The linear differential
equation is then treated straightforwardly (see Appendix
\ref{sec:annexe-semi-invasive}), and the profile $n(x)$ as well as
the dependence $I(d)$ may be calculated for any value of the source
strength. It is found that the result $I(d)/I(0)$ depends on the
parameter $\beta_{I}$ which measures the 'invasiveness' of the probe
and combines the thickness of the sensing layer $\delta$ and its
absorption coefficient $\gamma_{SL}$, as \
\begin{equation}
\beta_{I}=\frac{1-\kappa^{2}/\kappa_{SL}^{2}}{1+(\kappa/\kappa_{SL})\,coth\,\kappa_{SL}\delta}.
\label{eq:betaI-def}
\end{equation}
Here $\kappa$ and $\kappa_{S}$ stand, respectively, for
$\kappa\equiv(\gamma_{T}/D)^{1/2}=L_0^{-1}$ and
$\kappa_{SL}\equiv(\gamma_{SL}/D)^{1/2}$. The result for $I(d)$
reads as
\begin{equation}
\frac{I(d)}{I(0)}=\frac{1}{\cosh
(d/L_0)+\left(1-\beta_{I}\right)\sinh (d/L_0)}, \label{eq:I-si}
\end{equation}
with the limits of $\exp(-d/L_0)$ and $1/\cosh(d/L_0)$, respectively
for the case of non-invasive ($\beta_{I}\rightarrow0$), and strongly
invasive probe ($\beta_{I}\rightarrow1$). Fig. \ref{fig:nx-si} shows
that a thin layer efficient in exciton trapping and recombination
considerably affects the shape of $n(x)$, the distribution of
excitons in space. Related effect on the dependence of the signal
$I(d)$ on the distance from the source, described through Eq.
(\ref{eq:I-si}), is shown in Fig. \ref{fig:I-si}. Even for strongly
invasive probe, the effect shows mostly for $d<L_0$, while the
dependence $I(d)\propto\exp(-d/L_0)$ is restored at bigger
distances. \\
In Sec. \ref{sec:structure-optimisee}, the value of the diffusion
length was inferred from the experimental data without taking into
account the invasiveness of the sensing layer. According to the
analysis proposed in this part, the "real" diffusion length should
then be about 15-30 \% shorter than what plotted in Fig.
\ref{fig:long-eff}.\\

\section{Bulk carrier recombinations and microcavity effects}\label{sec:artefacts}

In thick diodes, more precisely when the active layer has a
thickness comparable or higher than  $\lambda/4n$ ($n$ being the
refractive index and $\lambda$ the wavelength in vacuum), the
amplitude of the optical field is modulated at the scale of the
diode and thus the assumption that the intensity emitted by the
probe layer only reflects the exciton density at a given point is
not generally valid. As shown in Eq. (\ref{eq:diff-simple}), this
microcavity effect happens as a modulation of the light emitted by
the probe layer. Although it also modulates the exciton diffusion
pattern, its influence is especially important for direct carrier
recombinations, since the latter occur with equal probability across
the whole thickness of the material under investigation. This has
been illustrated in Sec. \ref{sec:structure-optimisee} and is
discussed in more detail hereafter with a NPB/CBP/BCP trilayer
structure, commonly used in previous reports on exciton diffusion
measurements \cite{sun,zhou}. The occurrence of direct carrier
recombinations is linked to the NPB/CBP junction, which properties
are discussed in the second part of this section.

\subsection{Microcavity effects}\label{sec:champoptique}

\begin{figure}
\begin{center}
\includegraphics[width=1\linewidth]{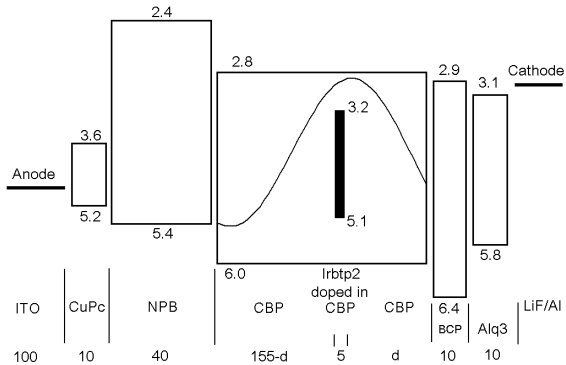}
\end{center}
\caption{Structure of the diode used in Sec. \ref{sec:champoptique}.
The layer thicknesses are indicated on the bottom in nanometers. The
HOMO and LUMO are taken from literature \cite{hill,ikai} and
specified for each compound (negative values). The profile of the
optical field is plotted in the CBP layer.}
    \label{fig:diodechampoptique}
\end{figure}

\begin{figure}
\begin{center}
\includegraphics[width=1\linewidth]{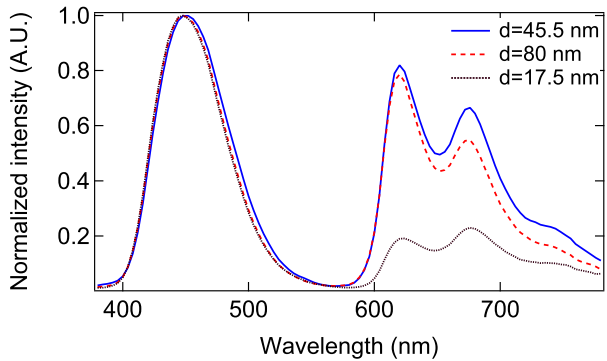}
\end{center}
\caption{Electroluminescent spectra of diodes described in Fig.
\ref{fig:diodechampoptique} for different $d$ positions at $J$=3.3
mA/cm$^2$.}
    \label{fig:spectre-champopt}
\end{figure}

\begin{figure}
\begin{center}
\includegraphics[width=0.9\linewidth]{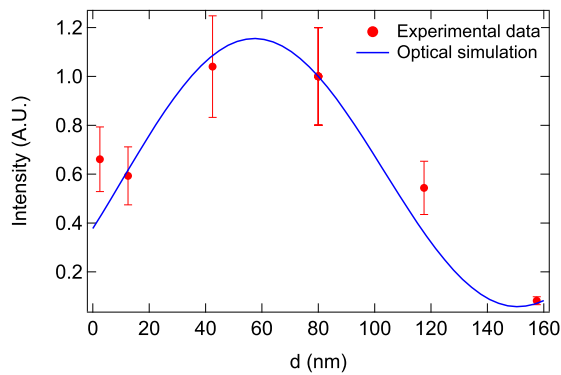}
\end{center}
\caption{The circles correspond to the red detected intensity
normalized to its blue part versus the $d$ position of the probe
layer at $J=3.3$ mA/cm$^2$. The profile of the optical field
calculated over the spectral range of Ir(btp)$_2$ emission is
plotted in continuous line.}
    \label{fig:resultat-champopt}
\end{figure}

Usual OLED devices are intrinsically weak microcavities, formed on
one side by a highly reflecting metallic cathode, and on the other
side by the ITO/Glass interface. As a consequence, there is a
stationary wave pattern inside the OLED, leading to a modulation of
the optical field. The probability to observe emission from a
molecule (either fluorescent or phosphorescent) depends on the
density of optical modes and on the effective mode volume at the
location of the molecule. The influence of the optical field
modulation has then to be taken into account to correctly describe
the OLED light emission. If this effect is not predominant in most
classical OLEDs because each layer thickness is only a few tens of
nanometers wide, it is clearly not the case in structures used for
diffusion length measurements where very thick layers are used
\cite{dandrade2,sun}. For instance, for the green emitter and the
structure used in \cite{sun}, the optical field is minimum
approximately in the middle of the layer and reaches a maximum near
the edges. Since expected diffusion patterns (exponential decays
from both edges) as well as optical field variations appear to go in
the same direction, it is likely that the extraction of  the only
diffusion effects in such a structure is problematic. To keep away
from this difficulty, we investigated a similar structure in which
the green emitter was replaced by a red one, in such a way that the
optical field was clearly distinguishable from a diffusion
signature, i.e. maximum in the middle of the CBP layer (see Fig.
\ref{fig:diodechampoptique}).\\
In order to calculate this emission probability, we assume that
every single molecule is an independent emitter creating its own
interference pattern. The planar geometry of the diode can be
considered as a Fabry-Perot type microcavity, characterized by a
strong dependence of the output light on both wavelength and
polarization \cite{jean,leger,benisty}. We here only consider the
light emitted at normal incidence, which is obviously not dependent
on polarization. The calculations were carried out using the ETFOS
software (Fluxim Inc.) which is based on the numerical method
descibed in \cite{etfos} and takes into account the
wavelength-dependence of the complex indices of all materials
involved in the full multilayer device. The model yields the
normalized intensity collected in the far-field as a function of
wavelength and position $d$ of the sensing layer.\\
The structure of the diodes is presented in Fig.
\ref{fig:diodechampoptique}. The diodes have been realized using the
experimental procedure described in Sec.
\ref{sec:structure-optimisee}. The spectra presented in Fig.
\ref{fig:spectre-champopt} show two contributions: one broad blue
peak due to NPB (see the comparison with photoluminescent spectra in
Fig. \ref{fig:pl}), and the characteristic structured spectrum of
Ir(btp)$_2$. In Fig. \ref{fig:resultat-champopt}, the red emission
(integrated over the spectral range of the phosphorescent emission),
plotted against $d$, is normalized with respect to the magnitude of
the blue NPB peak, to account for possible fluctuations of the total
luminance from one evaporation batch to another. A nice agreement is
noticed between the experimental data and the profile of the optical
field obtained from simulations \footnote{The only small
disagreement around $d=0$ can be explained by excitons diffusing
from the CBP/BCP interface as expected.}. The only adjusted
parameter here is the vertical scale.\\
This behavior could result from direct electron-holes recombinations
on the phosphorescent molecules or, alternatively, from the
diffusion of triplet excitons formed at the NPB/CBP and CBP/BCP
interfaces over a very large distance ($>$160 nm) so that the light
emission would eventually follow the shape of the optical field. To
discriminate between these two effects, we performed the same
experiment with DCM, a fluorescent compound, and we observed that
the measured red light versus the position $d$ agreed also very well
with the optical field (not shown here). As the DCM molecule could
not emit light from its triplet state and as singlet excitons can
not diffuse so far, the red light necessarily comes from
recombinations of carriers flowing in opposite directions in the
bulk. This suggests that the observed Ir(btp)$_2$ emission results
from direct exciton formation on these phosphorescent guests, which
is consistent with the hole trapping property of Ir phosphors
\cite{campbell}. Furthermore this observation implies that a
non-negligible amount of holes passes
through the NPB/CBP interface.\\
The example described in this Section provides a clear illustration
of a situation where microcavity effects dominate so much that no
diffusion length can be straightforwardly extracted from
experimental data. Although one may try to numerically factor-out
the contribution of the optical field from the experimental results,
the dominance of the microcavity effects implies that the
uncertainties, both on the experimental data and those related to
the optical field simulations (layer thicknesses, refractive
indices, etc.) reflect in the large uncertainties for the diffusion
length.

\subsection{Hole leakage through the NPB/CBP interface}\label{sec:npb/cbp}

\begin{figure}
\begin{center}
\includegraphics[width=1\linewidth]{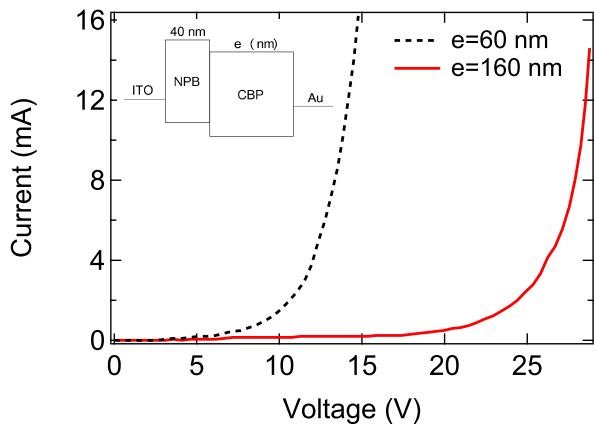}
\end{center}
\caption{Current density versus voltage curves for hole-only diodes
with different thicknesses of the CBP layer. Insert: structure of
the hole-only diodes.}
    \label{fig:iv-hole-only}
\end{figure}

\begin{figure}
\begin{center}
\includegraphics[width=0.8\linewidth]{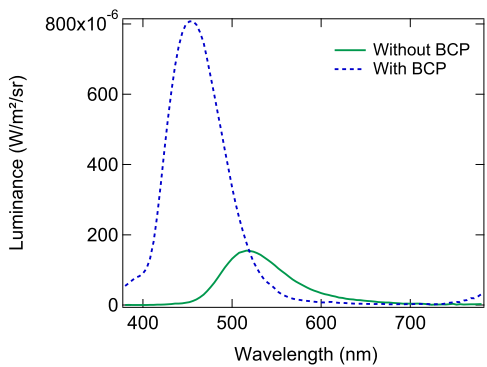}
\end{center}
\caption{Electroluminescent spectra of diodes NPB/CBP/Alq$_3$
(continuous line) and NPB/CBP/BCP/Alq$_3$ (dotted line) at $J$=1.7
mA/cm$^2$.}
    \label{fig:avec-sans-BCP}
\end{figure}

In the experiment described just above, the evidence of direct
recombinations in CBP also implies that holes do cross the NPB/CBP
interface. Here the heterojunction does not truly act as a blocking
interface for both types of carriers, and diffusion of excitons is
not the main channel of exciton generation in the sensing layer.
This feature has to be examined with some attention as it is in
contradiction with both basic energetic considerations and some
previous reports \cite{sun,zhou}.\\
Indeed, the NPB/CBP interface is usually described as an efficient
hole blocking barrier \cite{kim} due to the high energetic shift in
the HOMO levels of the two compounds \cite{hill}: the HOMO level of
NPB is higher than the HOMO level of CBP \cite{baldo2} by 0.5 eV
\cite{yang2} to 0.8 eV \cite{cheng}. However it is known that vacuum
levels may not be necessarily aligned at hetero-junctions, resulting
in an interface dipole. Although the interface dipole for NPB/CBP
has not been measured directly to the best of our knowledge, the
interface dipole in the case of organic-organic hetero-junctions is
determined by the difference in the charge neutrality levels
\cite{vasquez}, which have been measured independently and turn out
to be identical and equal to 4.2 eV \cite{vasquez}. As a result,
vacuum level alignment seems to be a valid assumption in this
case.\\
Yet, the experiments in Sec. \ref{sec:champoptique} show that holes
cross the barrier, which is not expected given that the energetic
jump (0.5 to 0.8 eV) is 20 to 30 times higher than thermal energy at
operating temperatures. According to OLED electrical models which
consider hetero-junctions, such as the MOLED code
\cite{tutis01,moled}, the hopping probability through such a high
energetic barrier is very unlikely, even when the effects of energy
disorder are taken into account
\cite{staudigel99,houli06a,houli06b}. It can be asked whether
electrons accumulated at the interface can help in some way the
leakage of holes. For instance, exciplexes could be suspected to
play the role of intermediate
states conveying holes from NPB to CBP.\\
To answer that question, electrons were removed by performing
"hole-only" experiments. Bilayer diodes have been fabricated,
consisting of NPB and CBP deposited between an ITO anode and a gold
cathode (see insert in Fig. \ref{fig:iv-hole-only}). The conditions
of deposition were similar to those described in Sec.
\ref{sec:structure-optimisee}, but the gold layer was only 20 nm
thick and the rate of deposition was less than 0.1 nm/s to limit the
likelihood of short-circuits which often appear with a gold cathode
due to the penetration of metallic particles into the organic
layers. The I-V curves for two different CBP layer thicknesses (60
and 160 nm respectively ; the NPB layer is kept equal to 40 nm) are
presented in Fig. \ref{fig:iv-hole-only}. They show a clear diode
behavior with a voltage threshold having the same order of magnitude
as the diodes described in Sec. \ref{sec:structure-optimisee}. In
addition, it can be seen that for a given current (ie. for a given
electric field in each material) the voltage drop across the whole
200-nm thick device is approximately two times the voltage drop
across the 100-nm thick device, at least well above threshold. Since
the two organic materials have comparable dielectric constants, this
probably indicates that the electric field is merely constant
throughout the device and then that the charge accumulated at the
junction is not significant (in contrast, a good hole-blocking
interface would yield a voltage drop ratio given by the ratio of the
CBP thicknesses, here $\sim$2.7). This evidences a hole flow through
the NPB/CBP interface in spite of the large difference between their
HOMO levels, without the assistance of electrons. Actually if
electrons would have been injected in the device, they would have
been detected through the blue emission of the NPB or CBP layer.
However, some diffusion of gold particles into the organic materials
cannot be completely ruled out, so that
percolation paths could be created for holes.\\
Another way to evidence that holes cross without the assistance of
electrons is to look at some exciton emission at the junction, which
is expected to come from NPB here since its gap is lower than the
CBP one. Indeed, from the electroluminescence spectra in Fig.
\ref{fig:avec-sans-BCP} of the NPB/CBP/BCP/Alq$_3$ diodes studied in
Sec. \ref{sec:champoptique}, we observed a blue peak assigned to
NPB, whereas, when the BCP layer was removed \footnote{The thickness
of the CBP layer was then 10 nm increased to minimize the changes in
the optical field.}, this NPB emission vanished while the
distinctive green emission spectrum of Alq$_3$ became clearly
visible. In this latter case, as the absence of NPB emission cannot
be attributed to microcavity effects, it confirms that very few
holes accumulate at the NPB/CBP interface, while most of them travel
up to the Alq$_3$ layer where they recombine with electrons. This
process cannot appear in the presence of the BCP layer, since the
CBP/BCP hetero-junction is a well-established hole blocking
interface, which is confirmed here. The appearance of NPB emission
in presence of BCP could be attributed to a reduced electric field
in CBP (resulting from the hole accumulation at the CBP/BCP
interface) which would make a little uneasy the hole crossing from
NPB to CBP.\\ This configuration is interesting since it almost
corresponds to a standard structure, studied by D'Andrade et al.
\cite{dandrade2}, and re-examined in more detail by Zhou et al.
\cite{zhou}, in which a thick Ir(ppy)$_3$-doped CBP layer was
inserted between NPB and pure CBP in order to infer the triplet
exciton diffusion length of CBP. In these works, all the excitons
were assumed to be created at the interface between NPB and
Ir(ppy)$_3$-doped CBP and subsequently diffuse in CBP  \cite{zhou}.
We believe that the efficiency saturation observed when increasing
the width of the doped layer in these devices cannot be attributed
to the single diffusion of CBP triplet excitons, even perturbed, or
shortened by the presence of the dopant. Firstly it could be argued
that given the energy gaps, as shown by the previous experiment,
excitons are formed rather in NPB, which might transfer their energy
to the phosphors via a Forster energy transfer in virtue of a good
spectral overlap. But since holes have the potential to easily cross
the barrier (as we have shown), it can be thought that some excitons
are formed directly on the phosphor, and then the generation zone of
excitons could be as large as a few nanometers, which matches the
value of 8 nm reported in \cite{dandrade2} for the exciton diffusion
length. Another important aspect that should be considered is the
opportunity for a Ir(ppy)$_3$ exciton to diffuse directly to another
guest site
\cite{ribierre,baldo}.\\
Through the lens of these experiments, it consequently appears that
the diffusion lengths obtained under the assumption of the NPB/CBP
hetero-junction being a highly efficient hole blocking barrier
\cite{dandrade2,zhou,sun} are questionable, and that a deeper
insight into this interface is needed to fully understand the
corresponding results.

\section{Conclusion}

In this work, we presented a thorough analysis of triplet exciton
diffusion length measurements by the repositionable thin sensing
layer technique in operational OLED devices. We demonstrated the
importance of a well-defined thin exciton generation zone to avoid
charge carriers flowing along the diode and thus leading to direct
recombinations in the probe layer. Through careful design of the
diode structure, we also circumvented masking effects of optical
field variations. As a result, a 16 $\pm$ 4 nm effective diffusion
length for triplet excitons in CBP was inferred from experimental
data in a working electroluminescent device. Comprehensive study of
the different processes at stake in such a structure allows
extraction of some useful guidelines for measuring triplet exciton
diffusion lengths in other materials:
\begin{enumerate}
\item An ideal case would be to create excitons only at the
heterojunction between the material of interest and a higher-gap
semiconductor, in order to avoid excitons being formed
preferentially in the latter. This is in practice challenging since
triplet host materials (especially for green or blue phosphors)
would be themselves high-bandgap materials. This was the case in the
illustrated example CBP here, which thus required the design of a
special structure.
\item In real cases, direct recombinations
occur, with a variable magnitude according to the injected current.
Away from the generation zone of excitons, the emission pattern
resulting from direct trapping of electrons and holes in the
repositionable thin sensing layer reproduce the shape of the optical
field inside the diode. The relative importance of direct
recombinations and diffusion can be straightforwardly measured by
replacing the phosphorescent dopant by a fluorescent one. In all
cases, the optical field modulates both the diffusion and direct
recombination pattern, and has to be carefully taken into account.
\item Triplet-triplet annihilation can be significant even at moderate
current densities and causes an apparent decrease of the measured
diffusion length. When the goal is to design a device where e.g.
color control is governed by exciton diffusion, it is important to
measure it at realistic current densities. If data with low
scattering are obtained, it is theoretically predicted that the
decay will not be a single exponential decay any more.
\item The choice of the probe layer (thickness and doping rate) should be
thought of as a trade-off between the intensity of light that can be
detected from low-exciton density regions, and its invasiveness, as
discussed in Section \ref{sec:th-layer}.
\end{enumerate}

\section*{Acknowledgments}

The authors are grateful to D. Tondelier and D. Ad\`es for
experimental and technological support and to E. Bogomolny and M.-C.
Castex for fruitful discussions. M. L. acknowledges the C'nano
Ile-de-France program for financial support. This work was supported
in part by Croatian MSES grant No.\ 035-0352826-2847.

\appendix

\section{Triplet-triplet quenching}\label{sec:annexe-TT}

In terms of rescaled, dimensionless quantities, $z=x/L_0$ and
$y(z)=n(x)/n_{0}$, the nonlinear equation acquires the form
\begin{equation} y''-y-y^{2}=0.\label{eq:app-1}\end{equation}
Multiplying this equation by $2y'$, the integration
gives\begin{equation}
y'^{2}-y^{2}-\frac{2}{3}y^{3}=C,\label{eq:app-2}\end{equation} $C$
being a constant. The equation is rewritten as (note the sign used
in taking the root),
\begin{equation}
\frac{dy}{\sqrt{y^{2}+\frac{2}{3}y^{3}+C}}=-dz.
\label{eq:app-3}
\end{equation}

In the case of \emph{non-invasive} probe, and very wide diffusion layer, the
requirement $n(x\rightarrow\infty)=0$ sets $C=0$. The integral on
the left hand side may then be expressed
in terms of hyperbolic functions. The dependence $y(z)$ (i.e.
$n(x)$) is then easily obtained and follows to Eq. (\ref{eq:nx-TT-ni-sinh}).\\

\section{Semi-invasive probe layer}\label{sec:annexe-semi-invasive}

Given the form of the differential equation for $n(x)$ (see Eq. (\ref{eq:DiffST})) in
the case without the triplet-triplet quenching, the solution is sought
in the form:
\begin{eqnarray}
Ae^{-\kappa x}+Be^{\kappa x}\hspace{0.5cm}&\textrm{for}&0<x<d\nonumber\\
Ce^{-\kappa_{SL}x}+Ee^{\kappa_{SL}x}\hspace{0.5cm}&\textrm{for}&d<x<d+\delta\nonumber\\
Fe^{-\kappa x}\hspace{0.5cm}&\textrm{for}&d+\delta<x\nonumber
\end{eqnarray}
Both $n(x)$ and its derivative should be continuous at junction
points, $x=d$ and $x=d+\delta$. Those requirements set the four
relations among coefficients $A,\  B,\  C,\  E$ and $F$. These,
together with the value of the strength $G$ of the source of
excitons at $x=0$, $G=-Dn'(0)=D\kappa\left(A-B\right)$, determine
the values of all the coefficients, $A$ to $F$. The calculation is
somewhat tedious if done manually, but straightforward if using some
of the usual software tools for symbolic computation. The limiting
case of strong invasiveness is particularly easy to treat.

\section{Invasive probe layer and triplet-triplet quenching}\label{sec:annexe-semi-invasive-and-TT}

\begin{figure}
\begin{center}
\includegraphics[width=1\columnwidth]{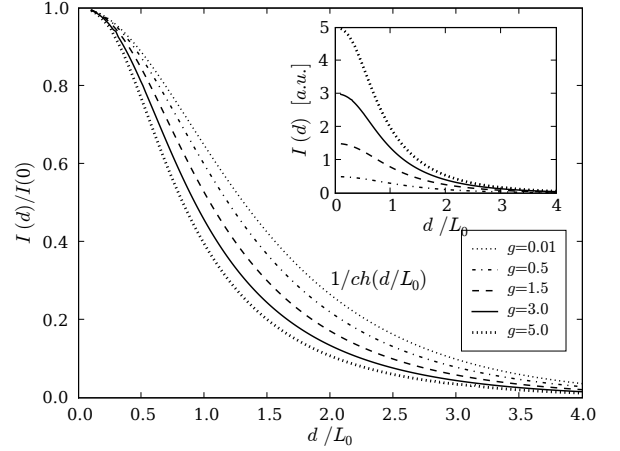}
\end{center}
\caption{The dependence of the strength of the signal $I(d)/I(0)$ on
the position $d$ of the probe for various strengths $g$ of the
triplet source (or triplet-triplet quenching rate). The insert shows
the graphs un-normalized by the strength $I(0)$ of the signal.}
    \label{fig:I-TT-inv}
\end{figure}

The nonlinear differential equation that poses in the presence of
triplet-triplet quenching may also be exactly solved in the presence
of infinitely invasive probe (i.e. the forcing $n(d)=0$). Then the
integration constant $C$ is directly related to the strength of the
signal $I(d)=-Dn'(d)\propto\sqrt{C}$ via Eq. (\ref{eq:app-2}).
Similarly, the strength of the source at $x=0$, $G=-Dn'(0)$, in
dimensionless quantities, is given by
\begin{equation}
g=\sqrt{y_{0}^{2}+\frac{2}{3}y_{0}^{3}+C}, \label{eq:app-4}
\end{equation}
where $y_{0}\equiv y(0).$ This is one of the equations to be used in
order to determine $C$ and $y_{0}$, while the second one is obtained
by integrating Eq. (\ref{eq:app-3}), \begin{equation}
\int_{0}^{y_{0}}\frac{dy}{\sqrt{y^{2}+\frac{2}{3}y^{3}+C}}=-\int_{d/L_0}^{0}dz=\frac{d}{L_0}.\label{eq:app-5}\end{equation}
This integral may be expressed in terms of the elliptic integral of
the first kind $F(\psi\backslash\alpha)$. From equations
(\ref{eq:app-4}) and (\ref{eq:app-5}), the values $y_{0}$ and $C$
for a given value $g$ and $d/L_0$ are calculated numerically (as
well as the value of the integral or related function
$F(\psi\backslash\alpha)$). The desired quantity $I(d)/I(0)$ is then
obtained from \begin{equation}
\frac{I(d)}{I(0)}=\frac{I(d)}{G}=\frac{\sqrt{C}}
{\sqrt{y_{0}^{2}+\frac{2}{3}y_{0}^{3}+C}}\leq1.\label{eq:app-6}\end{equation}
These results cannot be expressed in terms of elementary functions
and the curves are directly plotted in Fig. \ref{fig:I-TT-inv}. As
in the case without triplet-triplet quenching, the invasiveness of
the sensing layer is affecting the dependence of the signal at small
distances, in the same region where the triplet-triplet quenching
effect is strongest, and for strong exciton source, the dependence
at $d<L_0$ is again much steeper than $\exp(-d/L_0)$. The
exponential behavior determined by $L_0$ re-establishes at distances
further than $L_0$. The insert in Fig. \ref{fig:I-TT-inv} shows the
dependence of un-normalized signal for different strengths of the
source.


\end{document}